\documentclass[pre,twocolumn,showpacs,showkeys,superscriptaddress,floatfix]{revtex4}

\usepackage{graphicx}
\newcommand{\be}{\begin{equation}}
\newcommand{\ee}{\end{equation}}
\newcommand{\bea}{\begin{eqnarray}}
\newcommand{\eea}{\end{eqnarray}}

\begin{document}

 \title{Numerical Estimation of the Asymptotic Behaviour of Solid Partitions 
of an Integer}
 \author{Ville Mustonen}
 \affiliation{Department of Physics - Theoretical Physics, University of
Oxford, 1 Keble Road, Oxford OX1 3NP, UK}
 \affiliation{Helsinki University of Technology, Laboratory of Computational 
Engineering, P.O. Box 9203, FIN-02015, Finland}
 \author{R. Rajesh}
 \affiliation{Department of Physics - Theoretical Physics, University of
Oxford, 1 Keble Road, Oxford OX1 3NP, UK}
 \date{\today}
 \begin{abstract}

The number of solid partitions of a positive integer is an unsolved
problem in combinatorial number theory. In this paper, solid partitions
are studied numerically by the method of exact enumeration for integers
up to $50$ and by Monte Carlo simulations using Wang-Landau
sampling method for integers up to $8000$. It is shown that
$\lim_{n\rightarrow \infty} \frac{\ln\left(p_3(n)\right)}{n^{3/4}} 
= 1.79 \pm 0.01$, where $p_3(n)$ is the number of solid partitions of 
the integer $n$. This
result strongly suggests that the MacMahon conjecture for solid
partitions, though not exact, could still give the correct leading
asymptotic behaviour.
 \end{abstract}

\maketitle

\section{\label{sec1}Introduction}

Combinatorial enumeration problems arise naturally in many problems of
statistical physics. The number of partitions of an integer (see
\cite{andrews,stanley} for an introduction) is one such enumeration problem 
with a 
history dating back to Euler. Examples of applications to physical
problems include the $q\rightarrow \infty$ Potts model \cite{wurollet}, 
compact lattice animals \cite{wurollet,bhatia},
crystal growth \cite{temperley1}, lattice polygons \cite{rensburg}, 
Bose-Einstein statistics
\cite{holthaus,temperley2} and dimer coverings \cite{elser}.
The solution to the integer partitioning problem is known
for $1$-dimensional and $2$-dimensional partitions. However, not much 
is known about higher dimensional partitions. Numerical estimation of
the asymptotic behaviour of these higher dimensional partitions could lead 
to theoretical insights.
In this paper, we determine numerically the leading asymptotic behaviour of
$3$-dimensional partitions by exact enumeration and Monte Carlo
techniques.

\begin{figure} 
 \includegraphics[width=\columnwidth]{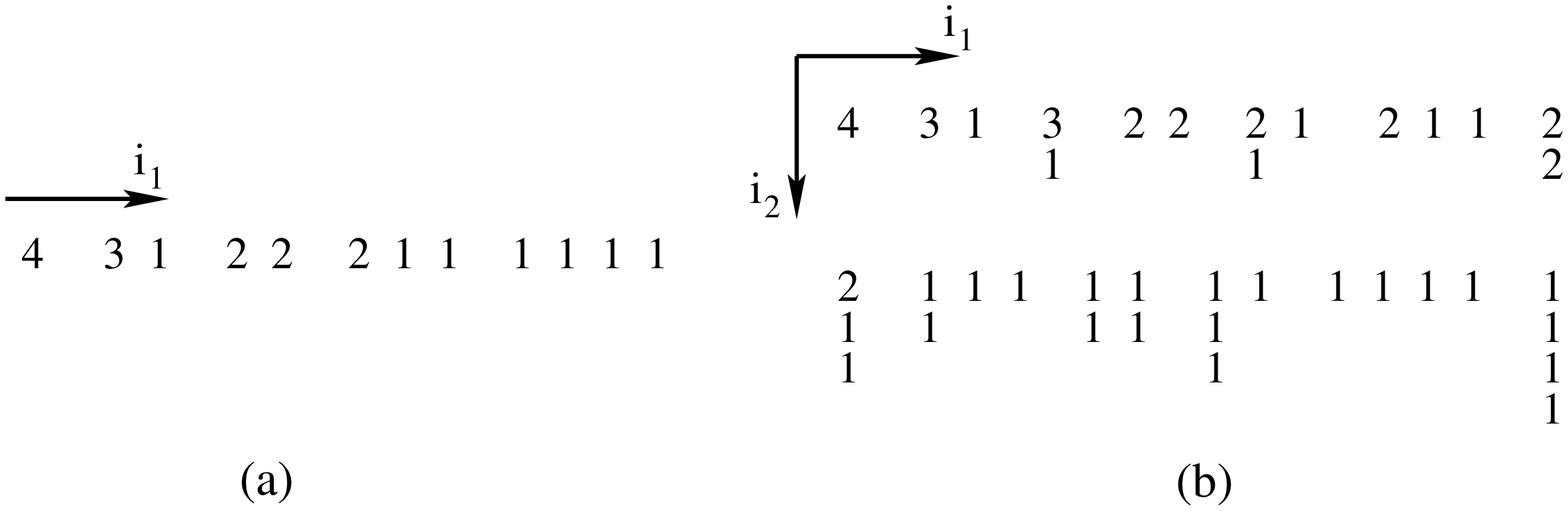}
 \caption{\label{fig4}Partitions of the integer $4$ in (a) one dimension
and (b) two dimensions.}
 \end{figure}

A $1$-dimensional or linear partition of an integer is a decomposition
into a sum of positive integers in which 
the summands are ordered from
largest to smallest.  A $2$-dimensional or plane partition of an integer
is a decomposition into a sum of smaller positive integers which are
arranged on a plane. The ordering property generalises to the summands
being non-increasing along both the rows and the columns.
Generalisation to $d$-dimension is straightforward.
Consider a $d$-dimensional hyper cubic lattice with sites labelled by ${\bf
i} = (i_1, i_2, \ldots, i_d)$, where $i_k = 1,2,\ldots$. An integer height
$h({\bf i})$ (corresponding to a summand)
is associated with site ${\bf i}$. A $d$-dimensional
partition of a positive integer $n$ is a configuration of heights such
that
 \bea
 h({\bf i}) & \geq & 0, \nonumber \\
 h({\bf i}) & \geq & \max_{1\leq k \leq d} ~h(i_1, i_2, \ldots, i_k +1, 
\ldots, i_d), 
\label{eq:definition} \\
 \sum_{{\bf i}} h({\bf i}) & = & n.  \nonumber
 \eea
 The second condition in Eq.~(\ref{eq:definition}) means that the heights
$h({\bf i})$ are non-increasing in each of the $d$ lattice directions.
As an illustration, the linear and plane partitions of
$4$ are shown in Fig.~\ref{fig4}(a) and \ref{fig4}(b) respectively.

Let $p_d(n)$ denote the number of partitions of $n$ in $d$-dimensions. 
The generating function $G_d(q)$ is then defined as
 \be
 G_d (q) = \sum_{n=0}^{\infty} p_d(n) q^n, 
 \ee
 where $p_d(0) \equiv 1$. 
The generating function for linear partitions is due
to Euler and is 
 \be
 G_{1}(q) = \prod_{k=1}^\infty (1-q^k)^{-1},
 \ee
 and $p_{1}(n)$ for large $n$ varies as \cite{hardyramanujan}
 \be
 p_{1}(n) \sim \frac{1}{4 n \sqrt{3}} \exp\left(\pi \sqrt{\frac{2 n}{3}}
\right), \quad n \gg 1. 
 \ee
 The corresponding formulae for plane partitions are
\cite{macmahon}
 \be 
 G_{2}(q) = \prod_{k=1}^\infty (1-q^k)^{-k},
 \label{2dmac}
 \ee
 and
 \be
 p_{2}(n) \sim \frac{c_2}{ n^{25/36}} \exp\left( \alpha_2 n^{2/3} \right),
\quad n \gg 1, 
 \ee
 where $c_2=0.40099\dots$ and $\alpha_2 = 2.00945\ldots$ \cite{wright1}. 
While the generating functions for three and
higher dimensional partitions are
not known, it is known that $\lim_{n\rightarrow \infty}
\ln \left(p_d(n)\right)/n^{d/(d+1)}$ has a finite
non-zero limit \cite{bhatia}. We define $\alpha_d$ to be 
 \be
 \alpha_d=\lim_{n \rightarrow \infty} \frac{\ln\left(p_d(n)\right)}
{n^{d/(d+1)}}.
 \ee
 In this paper, we numerically estimate $\alpha_3$ for solid 
($3$-dimensional) partitions to be
 \be
 \alpha_3 = 1.79 \pm 0.01.
 \label{eq:answer}
 \ee

Generalising the results for linear and plane partitions 
to higher dimensions, MacMahon suggested that
the generating function for $d$-dimensional partitions could be
\cite{macmahon}
 \be
 G^{(m)}_{d}(q) = \prod_{k=1}^\infty (1-q^k)^{- {k+d-2 \choose d-1}}. 
 \label{eq:macmahon}
 \ee
 Equation~(\ref{eq:macmahon}) is usually known as the MacMahon conjecture. 
Clearly, $G^{(m)}_d(q)$
is the correct result for $d=1,2$. However, it is known that $G^{(m)}_d(q)$ is
different from $G_d(q)$ for all $d \geq 3$ \cite{atkin,wright2}. In
particular, in $3$-dimensions
 \be
 G^{(m)}_{3}(q) = \prod_{k=1}^\infty (1-q^k)^{- k(k+1)/2},
 \label{eq:solid}
 \ee
 gives the wrong answer for the number of 
solid partitions of $6,7,8,\ldots$.

Let $p_d^{(m)}(n)$ denote the coefficient of $q^n$ in $G^{(m)}_{d}(q)$.
The asymptotic behaviour of $p_3^{(m)}(n)$ for large $n$
can be determined from Eq.~(\ref{eq:solid}) by the method of steepest
descent. In Appendix~\ref{appendix1}, we present a heuristic derivation
of the large $n$ behaviour of the coefficient of $q^n$ in the expansion
of the infinite product
 \be
 F(q) = \prod_{k=1}^{\infty} (1-q^k)^{-a_1 k^2 -a_2 k -a_3}.
 \label{eq:asymptotic}
 \ee
 Substituting $a_1=1/2$, $a_2=1/2$ and $a_3=0$ in Eq.~(\ref{eq:a5}), 
we obtain
 \be
 p^{(m)}_3(n)\sim \frac{c_3^{(m)}}{n^{61/96}} \exp\!\left(\alpha_3^{(m)}n^{3/4}
\!+ \!\beta_3^{(m)}n^{1/2} + \gamma_3^{(m)}n^{1/4} \right ), 
 \ee
 where $c_3^{(m)}$ is a constant and
 \bea
 \alpha_3^{(m)}&=& \frac{2^{7/4} \pi}{ 15^{1/4} 3} = 1.7898\ldots,
\label{eq:alpha}\\
 \beta_3^{(m)}&=& \frac{\sqrt{15} \zeta(3)}{\sqrt{2} \pi^2} =
0.3335\ldots,\\
 \gamma_3^{(m)}&=& -\frac{15^{5/4} \zeta(3)^2}{2^{7/4} \pi^5} =
-0.0414\ldots.
 \eea

Comparing the values for $\alpha_3^{(m)}$ in Eq.~(\ref{eq:answer}) and
$\alpha_3$ in Eq.~(\ref{eq:alpha}), we conclude that the MacMahon conjecture, 
though not exact, could still give the correct leading asymptotic 
behaviour for solid partitions. The value of $\alpha_3^{(m)}$ is a
function of only $a_1$ in Eq.~(\ref{eq:asymptotic}). Thus, if we assume
that the asymptotic behaviour for solid partitions is correctly captured
by a product form as in Eq.~(\ref{eq:asymptotic}), then it should have
the form $\prod_k (1-q^k)^{-(1/2\pm 0.012) k^2}$.

The rest of the paper is organised as follows. In Sec.~\ref{sec2}, we
present the results of the exact enumeration study. In Sec.~\ref{sec3},
we describe the Monte Carlo algorithm and the simulation results for
plane and solid partitions. 
Finally, we conclude with a summary and conclusions in Sec.~\ref{sec4}.

\section{\label{sec2}Exact enumeration}

Previous attempts at studying solid partitions on the computer have been
based on exact enumeration \cite{atkin,knuth,bratley,huangwu}. Tables of
$p_3(n)$ exist for $n$ up to $28$ \cite{knuth}. The table 
is extended up to $n=50$ in this paper by using
the standard back tracking algorithm \cite{martin}.
The algorithm is made faster by the
following. Partitions that are related to each other by symmetry
operations are counted only once and multiplied by the corresponding
symmetry factor. Also, parts of the partition that are
restricted to planes are generated by using the known generating
functions for plane partitions. In Table~\ref{table1},
we list the solid partitions from $n=29$ to $n=50$. For solid partitions
up to $n=28$, we refer to Ref.~\cite{knuth}.
\begin{table}
\caption{\label{table1} Solid partitions for $n=26$ to $n=50$. }
\begin{ruledtabular}
\begin{tabular}{lr|lr}
$n$ & $p_{3}(n)$ & $n$ & $p_{3}(n)$\\
\hline
 29 &         714399381  & 40 &      352245710866  \\
 30 &        1281403841  & 41 &      605538866862  \\
 31 &        2287986987  & 42 &     1037668522922  \\
 32 &        4067428375  & 43 &     1772700955975  \\
 33 &        7200210523  & 44 &     3019333854177  \\
 34 &       12693890803  & 45 &     5127694484375  \\
 35 &       22290727268  & 46 &     8683676638832  \\
 36 &       38993410516  & 47 &    14665233966068  \\
 37 &       67959010130  & 48 &    24700752691832  \\
 38 &      118016656268  & 49 &    41495176877972  \\
 39 &      204233654229  & 50 &    69531305679518  \\  
\end{tabular}
\end{ruledtabular}
\end{table}

\begin{figure} 
 \includegraphics[width=\columnwidth]{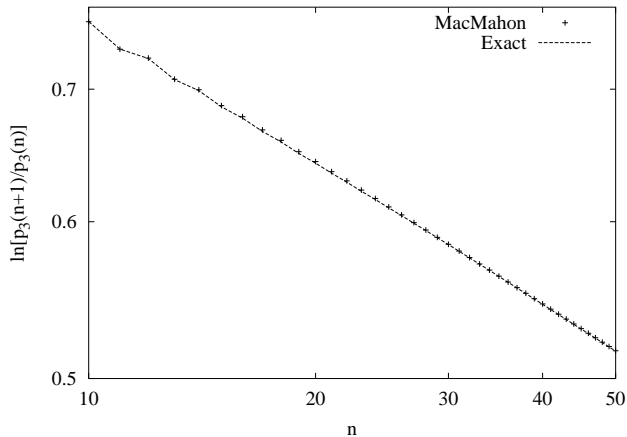}
 \caption{\label{fig1}The results from exact enumeration are compared
with $p_3^{(m)}$ obtained from the MacMahon conjecture.}
 \end{figure}

We compare the exact enumeration results with the answer predicted
by the MacMahon conjecture. In Fig.~\ref{fig1}, we show the 
variation of $\ln[p_3(n+1)/p_3(n)]$ with $n$ for both $p_3(n)$ as well
as $p_3^{(m)}(n)$. While there seems to be a good agreement, we are
unable to determine the precise asymptotic behaviour of $p_3(n)$
from these $50$ numbers. This is possibly due to the presence of
strong corrections to the leading asymptotic behaviour. It is
difficult to further extend the table of solid partitions due to the
large computing times involved. One possible method of probing larger
values of $n$ is to use Monte Carlo simulations. These are described in
Sec.~\ref{sec3}.

\section{\label{sec3}Monte Carlo simulation}

\subsection{\label{sec3a}Algorithm}

We use an algorithm proposed recently by Wang and Landau for measuring
density of states in spin systems \cite{landau}. The algorithm is
described below. Consider a $N_x\times N_y\times N_z$ lattice with 
initial height $h({\bf i})$ assigned to each lattice point 
in such a way that the configuration is a valid solid
partition.  To each positive integer $n$ is associated a histogram 
$H(n)$ and the number of solid partitions $p_3(n)$. The histogram $H(n)$
keeps track of the number of times solid partitions of $n$ have been
visited during the simulations.
The algorithm is based upon the fact that
if the probability of transition to a solid partition $n$ is proportional to
$[p_3(n)]^{-1}$, then a flat distribution is generated for the histogram
$H(n)$.  At the start of the program $H(n)=0$ and $p_3(n)=1$. 

A site is chosen randomly and as a trial move the height $h({\bf i})$ 
is increased or
decreased by one with equal probability, provided that the new state
is an allowed partition. If the new state is an allowed partition, then
the move is accepted with probability
 \be
 \mbox{Prob}(n_{old} \rightarrow n_{new}) = \min \left[\frac{p_3(n_{old})}
{p_3(n_{new})},1 \right],
\label{accProb}	
 \ee
 where $n_{old}$ and $n_{new}$ are the sum of heights for the old and
new states respectively, i.e, $n_{new}=n_{old}+1$, $n_{old}-1$ or
$n_{old}$ depending on
whether the height increased by one, decreased by one or remained the
same. The histogram $H(n)$ and $p_3(n)$ are updated as
 \bea
 H(n_{new}) &=& H(n_{new})+1, \label{update1}\\
 p_3(n_{new}) &=& f_{i} p_3(n_{new}),
\label{update2}
 \eea
 where $f_i$ is a modification factor greater than $1$. 

These steps are repeated until a flat histogram is created; in
practice, this means that $H(n)_{min} > c H(n)_{ave}$, where $c$ is a
flatness criteria typically between $0.75$ and $0.9$ while $H(n)_{min}$ is the
minimum of the $H(n)$'s and $H(n)_{ave}$ is the average of the
$H(n)$'s. When the histogram becomes flat, the modification factor $f_i$ is
changed to
 \be
 f_{i+1}=f_i^a,
 \ee
 and the histogram is reset to zero.
The exponent $a$ is less than $1$ and defines the smoothness of the
iteration. The program runs until $f$ is less than a predetermined value
$f_{final}$. 

Note that the algorithm does not obey detailed balance during the
simulations. However, in the 
limit $f_i \to 1$ when $p_3(n)$ takes its correct
value, the system does obey detailed balance with the weight of a solid
partition of $n$ being proportional to $[p_3(n)]^{-1}$.

The algorithm can be made faster by adopting certain ideas from the $N$-fold 
method \cite{binder,bortz}. In this modification, sites which cannot undergo
a valid move are never chosen. To do so, we define four classes.
(i) c1: sites at which the height can only increase.
(ii) c2: sites at which the height can either increase or decrease.
(iii) c3: sites at which the height can only decrease.
(iv) c4: an auxiliary class to help implementation of detailed balance.

First, we update the histogram $H(n)$ and $p_3(n)$ by
$\Delta$ times with $\mbox{Prob}(\Delta=k) = p'^k(1-p')$, where
$p'=(|c_4|+|c_1|/2+|c_3|/2)/(|c_1|+|c_2| +|c_3|+|c_4|)$ and $|c_k|$
denotes the number of elements in the class $c_k$.
We then choose one of the classes c1, c2, c3 with probabilities 
$|c_1|/(|c_1| + 2|c_2| +|c_3|)$, 
$2 |c_2|/(|c_1| + 2|c_2| +|c_3|)$ and 
$ |c_3|/(|c_1| + 2|c_2| +|c_3|)$ respectively.  
A site is picked up randomly from the chosen class and a trial move is
decided, for example if site ${\bf i}$ from class c1 is chosen, 
the trial move is $h({\bf i})=h({\bf i})+1$, while in the  case of class
c2 the height increases or decreases with equal probability. Finally,
we either accept or reject the trial move according to Eq.~(\ref{accProb}) and
update the histogram and $p_3(n)$ according to Eqs.~(\ref{update1}) and
(\ref{update2}).
With this construction, only valid trial moves are chosen at each time 
step and the algorithm becomes considerably faster. 
The role of the class c4 is to make the algorithm obey detailed balance 
asymptotically, i.e. when
$f_{i} \to 1$. 
We define $|c4|=C-(|c_1|+|c_2|+|c_3|)$, where $C$ is some large enough 
constant. 

Further speeding up can be done by dividing the interval 1--8000 to
smaller slightly overlapping intervals ($14$ in our case) 
and the simulation is done for each interval.
After the simulations are over, these intervals can be 
joined together to produce $p_3(n)$. The distribution is finally
normalised by setting $p_3(1)=1$.

The parameters we have used for the simulations are 
$c=0.85$, $a=1.0/1.4$, $C=2000$, $f_0= 2.5 $ and 
$f_{final}=1.0000099$ (corresponding to $35$ iterations).  Changing
these parameters slightly does not change the final outcome of the
simulation.
The lattice sizes used for plane and solid
partitions were $100\times 100\times 1$ and $50\times 50\times 50$
respectively. 
Random numbers were generated using standard RANMAR algorithm.  With
this setup, a typical run producing one $p_{3}(n)$ for $n$ between
1--8000 takes about 12
hours with a Pentium 4 processor. For statistics we performed 20 runs for
plane partitions and 24 runs for solid partitions using different random
number sequences. Since $p_{3}(n)$ is typically a very large number, the
quantity that we keep track of in the simulations is $\ln(p_{3}(n))$. 

\subsection{\label{sec3b}Simulation results for plane partitions}

\begin{figure} 
 \includegraphics[width=\columnwidth]{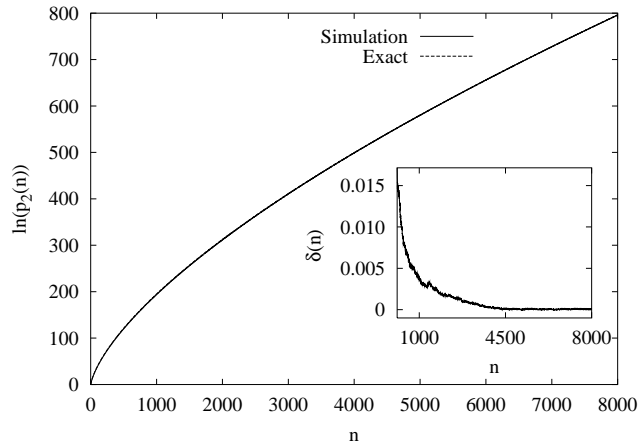}
 \caption{\label{fig2}The simulation results for plane partitions are
compared with the exact answer. In the inset, the variation of the
relative error with $n$ is shown.}
 \end{figure}

We first test the algorithm against the known case of plane partitions.
In Fig.~\ref{fig2}, we compare the simulation results with the exact
answer. The two curves are almost indistinguishable. Consider the
relative error defined by
 \be
 \delta(n)=\frac{|\ln(p_{2}^{(s)}(n))-\ln(p_{2}(n))|}{\ln(p_{2}(n))},
 \ee
 where $p_2^{(s)}(n)$ is the value obtained from simulations. We show
the variation of $\delta(n)$ with $n$ in the inset of Fig.~\ref{fig2}.
The relative error goes to zero for large $n$. Thus, we conclude
that the algorithm does give the correct leading asymptotic behaviour.
In principle, the simulations can be made arbitrarily precise, and the
correction terms can be determined. However, in our case, the
statistical errors are not small enough to allow a reliable
determination of the correction terms to the leading asymptotic
behaviour.

\subsection{\label{sec3c}Simulation results for solid partitions} 

For solid partitions, we calculated $p_3(n)$ numerically by averaging
over $24$ different runs. In Fig.~4, we show the results from simulation
while in the inset of Fig.~4, we show the relative error. We estimate
the asymptotic behaviour by fitting the data to an assumed form by the
method of least squares fit. We fit $\ln[p_3^{(s)}(n+1)/p_3^{(s)}(n)]$ in the
range 20--8000 to the form $0.75 \alpha_3 n^{-1/4} + 0.5 \beta_3 n^{-1/2}
+ 0.25 \gamma_3 n^{-3/4}+d_3 n^{-1}$, where $p_3^{(s)}(n)$ is the value
for solid partitions obtained from simulations.
We choose this form since the
MacMahon conjecture has the same functional form. As a test for the
fitting routine, we test it against the simulation results as well as 
against the exact
results for plane partitions using the fitting form $0.667 \alpha_2
n^{-1/3} + 0.333 \beta_2 n^{-2/3}
+d_2 n^{-1}$. The results for $\alpha_{(2,3)}$ are presented in
Table.~\ref{table2}.

\begin{figure} 
 \includegraphics[width=\columnwidth]{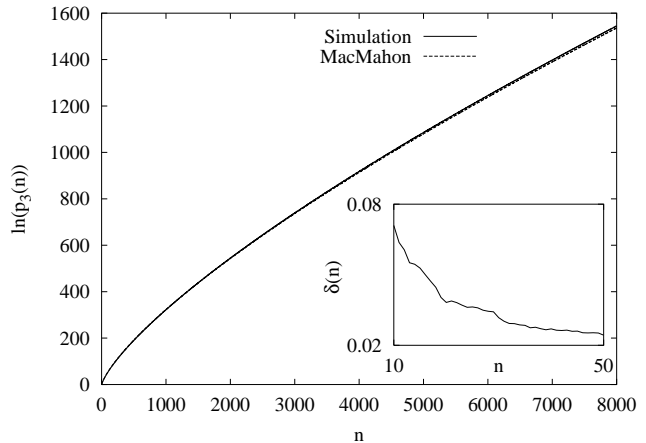}
 \caption{\label{fig3}Simulation results for solid partitions are
compared with $p_3^{(m)}$ obtained from the MacMahon conjecture. In the
inset, we show  the relative error with respect to the answer obtained 
from exact enumeration.}
 \end{figure}
\begin{table}
\caption{\label{table2} The results obtained from the least square fit
are shown. $\alpha_2$, $\alpha_2^{(s)}$, $\alpha_3^{(m)}$ and
$\alpha_3^{(s)}$ correspond to $p_{(2,3)} (n)$ obtained from the exact
results for plane partitions Eq.~(\ref{2dmac}), Monte Carlo results for
plane partitions, MacMahon conjecture for solid partitions
Eq.~(\ref{eq:solid}) and Monte Carlo results for solid partitions
respectively.}
\begin{ruledtabular}
\begin{tabular}{cccc}
$\alpha_2$ & $\alpha_2^{(s)}$ & $\alpha_3^{(m)}$ & $\alpha_3^{(s)}$\\
$ 2.010 \pm 0.002$ &  $ 2.01\pm 0.01$ &$ 1.789\pm 0.002$ & $ 1.79 \pm 0.01$ \\
\end{tabular}
\end{ruledtabular}
\end{table}

For plane partitions, the value of $\alpha_2$ obtained from the fitting
routine is in excellent agreement with the exact answer 2.00945..
\cite{wright1}. Agreement with the correct answer is also seen for
$\alpha_3^{(m)}$ (see Eq.~(\ref{eq:alpha})). Hence, we conclude that
$\alpha_3 = 1.79 \pm 0.01$.

\section{\label{sec4}Summary and conclusions}

In summary, we studied numerically the problem of solid partitions of an
integer. Using exact enumeration methods, we extended the table of solid
partitions for integers up to $50$. However, we were unable to determine
the precise asymptotic behaviour of solid partitions from these $50$
numbers. Solid partitions for larger values of $n$ were studied using
Monte Carlo simulations.  From these simulations, we showed
that $\lim_{n\rightarrow \infty} n^{-3/4} \ln\left(p_3(n)\right) =1.79 
\pm 0.01$.
This value is consistent with the MacMahon value 
for solid partitions. 
Thus, if we assume
that the asymptotic behaviour for solid partitions is correctly captured
by a product form as in Eq.~(\ref{eq:asymptotic}), 
then it should have
the form $\prod_k (1-q^k)^{-(1/2\pm 0.012) k^2}$.

\section*{Acknowledgements}

We would like to thank D.~Dhar, D.~B.~Abraham and R.~Stinchcombe 
for useful discussions.
VM was partially supported by the Academy of Finland, 
Research Centre for Computational Science and Engineering, project
No. 44897 (Finnish Centre of Excellence Programme 2000-2005), 
Wihuri Foundation, Finnish Cultural Foundation and Tekniikan 
edist\"amiss\"a\"ati\"o.
RR would like to acknowledge the financial support of EPSRC, UK.

\appendix

\section{\label{appendix1}Asymptotics for the MacMahon conjecture}

In this appendix, we present a heuristic derivation of the asymptotic
behaviour of the coefficient of $q^n$ in the expansion of the product
 \be
 F(q)= \prod_{k=1}^{\infty} (1-q^k)^{-a_1 k^2 - a_2 k -a_3}.
 \label{eq:a1}
 \ee
Let $q=e^{-\epsilon}$. Taking logarithms on both sides of Eq.~(\ref{eq:a1})
and converting the resulting summation into an integral by using the
Euler-Maclaurin summation formula (for example, see \cite{bender}), 
we obtain
 \bea
 \ln  \left( F(e^{-\epsilon})\right) &= &
\frac{2 a_1 \zeta(4)}{\epsilon^3}
+\frac{a_2 \zeta(3)}{\epsilon^2}
+\frac{a_3 \zeta(2)}{\epsilon} \nonumber\\
&&\mbox{} + \frac{a_2 + 6 a_3}{12} \ln(\epsilon) + O(\epsilon^0),
 \eea
 where $\zeta(n)$ is the Riemann zeta function.
Let the coefficient of $q^n$ in $F(q)$ be denoted by $c(n)$. Then
 \be
 c(n) = \frac{1}{2 \pi i}\oint \frac{F(q)}{q^{n+1}}.
 \ee
 For large $n$, we evaluate $c(n)$ by the method of steepest descent. 
The saddle point is the maximum of  $\epsilon n + \ln(F(\epsilon))$. 
This occurs at $\epsilon_0$, where
 \bea
 \epsilon_0 &=& \frac{a_1^{1/4} \pi}{15^{1/4}} n^{-1/4} + \frac{\sqrt{15}
 a_2 \zeta(3)}{2 \sqrt{a_1} \pi^2} n^{-1/2} \nonumber \\
&+& \!\! \frac{15^{1/4} (a_1 a_3
 \pi^6 - 45 a_2^2 \zeta(3)^2)}{24 a_1^{5/4} \pi^5} n^{-3/4} + 
O(n^{-1}).
 \eea
 Evaluating the integral about this saddle point, we obtain
 \bea
 \lefteqn{ \ln[c(n)] = 
\frac{4 a_1^{1/4} \pi}{15^{1/4} 3}n^{3/4} }
\nonumber \\
&& \mbox{} + \frac{\sqrt{15} a_2 \zeta(3)}{\sqrt{a_1} \pi^2} n^{1/2}
+ \frac{5^{1/4} (a_1 a_3 \pi^6 -
45 a_2^2 \zeta(3)^2)}{2 a_1^{5/4} 3^{3/4} \pi^5} n^{1/4} \nonumber \\
&& \mbox{} - \left(\frac{5}{8}+\frac{a_2}{48}+\frac{a_3}{8} \right) \ln(n) 
+ O(n^0).
 \label{eq:a5}
 \eea

\end{document}